\documentclass{aa}  

\usepackage{graphicx}
\usepackage{txfonts}
\newcommand{\doce}{\mbox{$^{12}$CO}}

\newcommand{\trece}{\mbox{$^{13}$CO}}

\newcommand{\jtd}{\mbox{$J$=3$-$2}}

\newcommand{\kms}{\mbox{km\,s$^{-1}$}}

\newcommand{\ms}{\mbox{$M_{\mbox{\sun}}$}}
\newcommand{\ls}{\mbox{$L_{\mbox{\sun}}$}}

\newcommand{\gsim}{\raisebox{-.4ex}{$\stackrel{\sf >}{\scriptstyle\sf \sim}$}}

\newcommand{\sou}{\mbox{IRAS\,08544$-$4431}}

\begin{document}

   \title{High-resolution observations of  \sou}
  
   \subtitle{Detection of a disk orbiting a post-AGB star and of a slow  disk wind} 

   \titlerunning{ALMA observations of \sou}
   
   \author{V. Bujarrabal
          \inst{1}
          \and
          A. Castro-Carrizo\inst{2} \and H. Van
          Winckel\inst{3} 
          \and J. Alcolea\inst{4} \and C. S\'anchez Contreras\inst{5} \and
M.\ Santander-Garc\'{i}a\inst{4,6} \and M.\ Hillen\inst{3} }

   \institute{             Observatorio Astron\'omico Nacional (OAN-IGN),
              Apartado 112, E-28803 Alcal\'a de Henares, Spain\\
              \email{v.bujarrabal@oan.es}
\and 
 Institut de Radioastronomie Millim\'etrique, 300 rue de la Piscine,
 38406, Saint Martin d'H\`eres, France  
\and
Instituut voor Sterrenkunde, K.U.Leuven, Celestijnenlaan 200B, 3001
Leuven, Belgium
\and
             Observatorio Astron\'omico Nacional (OAN-IGN),
             C/ Alfonso XII, 3, E-28014 Madrid, Spain
\and
Centro de Astrobiolog\'{\i}a (CSIC-INTA), Ctra. M-108, km. 4,
E-28850 Torrej\'on de Ardoz, Madrid, Spain 
\and 
Instituto de Ciencia de Materiales de Madrid (CSIC). Calle
             Sor Juana In\'es de la Cruz 3, E-28049 Cantoblanco, Madrid,
             Spain 
   }

   \date{Received 5 December 2017 / Accepted 23 January 2018}

  \abstract
   {}
{In order to study the effects of rotating disks in the post-asymptotic
  giant branch (post-AGB) evolution,
  we observe a class of binary post-AGB stars that seem to be
  systematically surrounded by equatorial disks and slow outflows.
  Although the rotating dynamics had only been well identified in three
  cases, the study of such structures is thought to be fundamental to
  the understanding of the formation of disks in various phases of the late
  evolution of binary stars and the ejection of planetary nebulae from
  evolved stars.  }
{We present ALMA maps of \doce\ and
  \trece\ \jtd\ lines in the source \sou, which belongs to the above
mentioned class of objects. We analyzed the data by means of nebula
  models, which account for the expectedly composite source and can
  reproduce the data. From our modeling, we  estimated the main nebula parameters, 
  including the structure and dynamics and the density and temperature
  distributions. We discuss the uncertainties of the derived values and, in
  particular, their dependence on the distance.}
{Our observations reveal the presence of an equatorial disk in
  rotation; a low-velocity outflow is also found, probably formed of
  gas expelled from the disk. The main characteristics of our
  observations and modeling of \sou\ are similar to those of
  better studied objects, confirming our interpretation.  The disk
  rotation indicates a total central mass of about 1.8 \ms,  for a
  distance of 1100 pc.  The disk is found to
  be relatively extended and has a typical diameter of $\sim$ 4 10$^{16}$ cm.
The total nebular mass is $\sim$ 2 10$^{-2}$
  \ms, of which $\sim$ 90\% corresponds to the
  disk. Assuming that the outflow is due to mass loss from the disk, we
  derive a disk lifetime of $\sim$ 10000 yr. The disk angular
  momentum is found to be comparable to that of the binary system at
  present. Assuming that the disk
  angular momentum was transferred from the binary system, as expected,
  the high values of the disk angular momentum in this and other similar
  disks suggest that the size of the stellar orbits has significantly
  decreased as a consequence of disk formation.
}
   {}


   \maketitle
%

\section{Introduction}

Planetary and preplanetary nebulae (PNe, pPNe) often show a remarkable
axial symmetry and fast bipolar outflows with velocities around 100
\kms. On the contrary, asymptotic giant branch (AGB) circumstellar
envelopes, their immediate precursors, are spherical, at least at
large-scale, and expand isotropically at moderate velocities (10 -- 20
\kms).
The development of
such axial structure and dynamics remains an open question. It has been
proposed to be associated with rotating disks (e.g., Soker 2001,
Bujarrabal et al.\ 2001, Balick \& Franck 2002, S\'anchez Contreras et
al.\ 2002), from which material would fall onto the star or a companion
during early post-AGB phases, powering very fast and collimated stellar
jets.  In principle, the material ejected during the AGB phases  does not
have enough angular momentum to form Keplerian disks, which should only 
appear around binary stellar systems, as these systems have the
necessary angular momentum stored in their orbital movement.  Other
mechanisms to explain bipolar post-AGB nebula involve an anisotropic
sudden ejection of stellar gas by a very late AGB (or very early
post-AGB) star during a common-envelope phase (Alcolea et al.\ 2007, De
Marco et al.\ 2009, 2011, Iaconi et al.\ 2017). However, our
theoretical understanding of these phenomena is still poor.

The detailed observation of Keplerian disks around post-AGB stars,
including their dynamics, is not straightforward, since it requires
high angular and spectral resolutions. To date, disks have been well
mapped only in three objects, the Red Rectangle, AC Her, and IW Car, by
means of interferometric mm-wave maps of CO lines (Bujarrabal et
al.\ 2013b, 2015, 2017).  These sources belong to a class of binary
post-AGB stars surrounded by low-mass nebulae that show independent
evidence of the presence of disks (e.g., Van Winckel 2003, de Ruyter
et al.\ 2006, Gezer et al.\ 2015, Hillen et al.\ 2016, 2017).
Notably, they are characterized by spectral energy
distributions (SEDs) with a significant near-infrared (NIR) excess, revealing the
existence hot dust
close to the stellar system. The very compact nature of the NIR
emission has been confirmed by interferometric IR measurements
(e.g.,\ Hillen et al. 2017).  In addition, the IR spectra reveal the
presence of highly processed grains, which is indicative of the
longevity of the disks.

Single-dish observations of \doce\ and \trece\ emission in these
NIR-excess post-AGB stars (including the Red Rectangle, AC Her, and IW
Car) systematically yielded characteristic line profiles, which are
exactly those expected from relatively extended Keplerian disks
(Bujarrabal et al.\ 2013a, Bujarrabal \& Alcolea 2013).  A slowly
expanding component was also proposed in most nebulae, 
  although its shape and dynamics could not be well identified from
  those data. Indeed, ALMA
maps of CO lines in the Red Rectangle and IW Car show a bipolar
low-velocity outflow, which is particularly noticeable in lines with relatively
high opacity and excitation; this bipolar outflow was deduced to be very probably
formed of gas extracted from the disk, given its structure and
kinematics.  The  CO emission image of 89 Her, another NIR-excess
post-AGB,  is dominated by an extended hourglass-like nebula in slow
expansion  (Bujarrabal et
al.\ 2007). Although rotation was not actually resolved in 89 Her, a
small disk could be confined to the prominent central clump.
Recent NOEMA maps of the similar objects IRAS\,19125+0343 and R Sct
(G\'omez-Garrido et al., in preparation) also show an expanding
nebula and a possible central disk that remains unresolved. On the
other hand, no gas in expansion was found in the mm-wave maps of AC Her,
but we cannot rule out a diffuse outflow, since the CO interferometric
data show a significant flux loss and no  sub-mm observations exist.

We stress that no signs of Keplerian disks have been found in other
kinds of post-AGB nebulae, in particular in the well-observed high-mass
pPNe and young PNe. We cannot exclude that confusion with the strong
emission from their expanding nebulae prevents in some way the
detection of emission from small central disks, but even
high-resolution observations of the very inner nebular regions have
yielded no sign of disks to date  (e.g.,\ Alcolea et al.\ 2007,
  Olofsson et al.\ 2015, Santander-Garc\'{\i}a et al.\ 2017).  The
  presence of binary systems in the center of high-mass bipolar pPNe
  and PNe is also debated. Some well-known nebulae,
   such as OH\,231.8+4.2, M\,2-9, and several evolved PNe,  harbor binary
systems (see,
e.g.,\ S\'anchez Contreras et al.\ 2004, Castro-Carrizo et al.\ 2012, 
Hillwig et al.\ 2016).  However, long-term radial velocity studies
  have yielded negative results in a number of post-AGB sources; for example,
  Hrivnak et al.\ (2017)
  have only found that one  object, out of seven well-studied sources,
  was probably a wide stellar system.
  These findings imply that there are
  significant constraints to eventual binarity in these sources.

 The evolution of NIR-excess post-AGB objects is not well known
  (e.g.,\ Van Winckel et al.\ 2009, De Marco 2014) and could be
  significantly different from that of well-studied (pre)planetary
  nebulae. Their nebular mass, including rotating and expanding gas,
is low, $<$ 0.1 \ms, often $\sim$ 0.01 \ms\ (Bujarrabal et
al.\ 2013a). This suggests that they are not ejecting sufficient mass to
form a high-mass PN (i.e.,\ a nebula containing most initial
mass). However, we point out that in many well-known PNe and pPNe the
total detected nebular mass is smaller than $\sim$ 0.1 \ms, including
ionized gas, molecular gas, or PDR-like components; see, e.g.,\ the
compilation of mass values by Pottasch (1984), Huggins \& Healey
(1989), Huggins et al.\ (1996), S\'anchez-Contreras et al.\ (2012), and
Castro-Carrizo et al.\ (2001).  It is also probable that the
interaction of the star with the orbiting disk, including reaccretion
of material, slows down their post-AGB evolution (e.g.,\ Van Winckel et
al.\ 2009). Indeed, all these NIR-excess stars still show relatively
low stellar temperatures and exhibit spectral types usually in the range
F-K. It is obvious that the  conspicuous nebulae detected in some
of these objects could form a low-mass PN, but the
star would not necessarily become hot enough to ionize the surrounding gas before, due
to expansion, the nebula becomes too diffuse to be detectable.

Most remarkably, disks are observed in binary post-AGB stars, which
often show orbits that are too small to accommodate an AGB star (Van
Winckel et al.\ 2009, Manick et al., 2017).  In the best-studied
NIR-excess post-AGB star, the Red Rectangle (Bujarrabal et al.\ 2016),
the total angular momentum of the disk is not negligible in comparison
to that of the binary at present. If, as expected, the disk angular
momentum originates from the stellar system during a previous phase of
strong interaction, a comparison of both the disk and binary momentum
values at present implies that a significant decrease of the distance
between the stars occurs (by a factor \gsim\ 2) from disk formation. We
reach a similar conclusion from our observations; see detailed
discussion in Sect.\ 4. These results suggest that the binary orbit was
wider than the size of an AGB star in the past; but not much wider,
which helps to explain the transfer of angular momentum.  A better
knowledge of the main properties of the disks is therefore imperative
to understand the orbits of evolved binary stellar systems.

Finally, we recall that our post-AGB stars are not the only evolved
stars surrounded by Keplerian disks.  The existence of disks
  orbiting white dwarfs (WDs) has been known for more than three
  decades
  (e.g.,\ Zuckerman
  \& Becklin 1987, Bil\'{\i}kov\'a et al.\ 2012). Some of these stars
  are also surrounded by prominent PNe, which are much more extended
  than the disks. It has been argued, see Clayton et al.\ (2014), that
  the Keplerian disks found in PNe are remnants of disks formed from
  circumstellar material ejected by the star in AGB or early post-AGB
  phases;  this process is very similar to that probably responsible for the
  disks we are studying. The disks around very old WDs can be
  different; these disks possibly originate from the disruption
  of asteroids or analogs. 
Moreover, the presence of Keplerian rotation in a certain class of AGB
stars, that is, semiregular variables that show aspherical shells in anisotropic
slow expansion, has been also proposed. A relatively small disk
  has been well identified in one of these objects, L$_2$ Pup (Homan et
al.\ 2017). Our sources could therefore represent an evolutionary link
between disks around AGB stars and disks around WDs  (at least
  those surrounded by PNe), through the phase of prominent post-AGB
disks. If this relation really holds, the disks surrounding WDs, whose
mass is very low, would just contain a small fraction of the previous
post-AGB disks, after a long process of disk evaporation.

We think that the study of disks orbiting NIR-excess post-AGB stars may
be fundamental to the understanding of the formation of disks in
various phases of the late evolution of intermediate-mass binary
stars. This phenomenon can be crucial to understanding the late
evolution of binary systems and the shaping of PNe.


We present ALMA maps of \sou\ in CO emission that clearly show both
rotating and expanding gas. \sou\ is a low-amplitude pulsator that
belongs to the class of NIR-excess post-AGB stars mentioned before
(Maas et al.\ 2003, De Ruyter et al.\ 2006, Bujarrabal et al.\ 2013a,
Hillen et al.\ 2016). The hot-dust component of \sou\ has been observed
in the IR using the VLTI. The inner rim of the disk was imaged, showing
a diameter of about 15 mas and suggesting an inclination with respect
to the plane of the sky of about 20$^\circ$.  Our CO data
confirm the disk-like structure and the value of the inclination,
although the size of the CO-emitting region is much larger.  \sou\ is a
double stellar system; see Van Winckel et al.\ (2009). Assuming that
the inclination of the orbit is that of the disk and a high-luminosity
post-AGB primary with $\sim$ 0.5 \ms\ at present, one deduces a $\sim$
1$-$2-AU-wide orbit and a secondary with about 1.3 \ms;  but we
  stress that the mass of the individual stars is not well determined
  and strongly depends on the orbit inclination (see further discussion
  in Sect.\ 3.1).
  
These authors suggested a distance $D$ $\sim$ 550 pc for \sou. However,
this value is based on an assumed standard luminosity of 3000 \ls\ and
is therefore very tentative. The GAIA parallax, 0.86 $\pm$ 0.6 mas, is
not very accurate and we recall that the binary nature of the star can
affect the parallax measurements in a complex way (Acke et
al.\ 2013). As we see later in this work, $D$ $\sim$ 1100 pc is more
compatible with our estimates of the total stellar mass, and we have
adopted this value (see Sect.\ 3.1). We widely discuss the effects of
the distance uncertainty on our modeling, in particular, to allow a
comparison with previous results.

\begin{figure*}
  
   \hspace{.3cm}
   \includegraphics[width=17.cm]{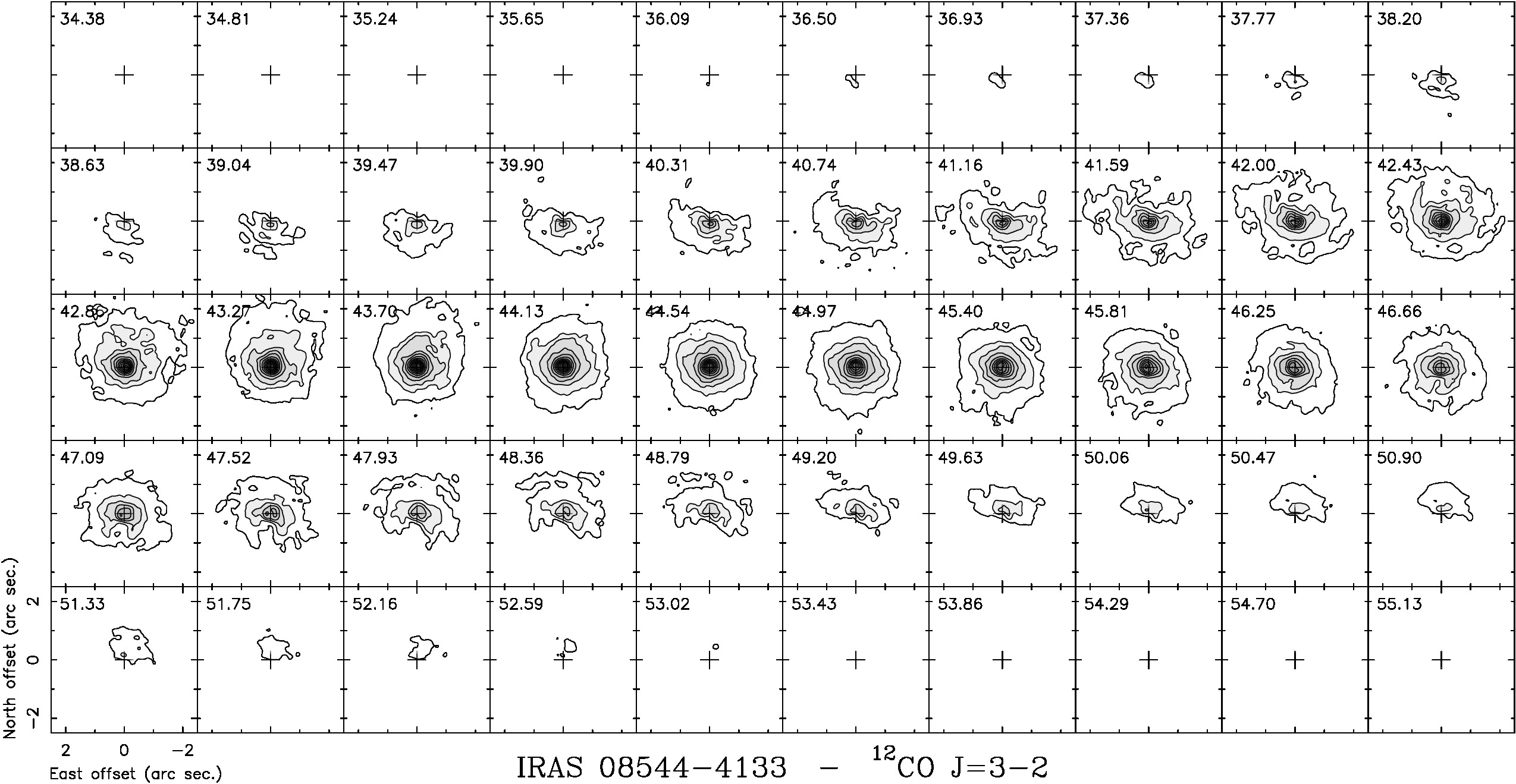}
    \caption{ALMA maps per velocity channel of \doce\ \jtd\ emission
      in \sou. The continuum emission has been subtracted to better show the
        distribution of the weak line. The first contour and spacing
        are 0.02 Jy beam$^{-1}$ (equivalent to 6 K, Rayleigh-Jeans equivalent
        temperature). The {\em LSR} velocities are indicated in each panel
        (upper left corner).  }
         \label{}
\end{figure*}

\begin{figure*}
  
   \hspace{.3cm}
     \includegraphics[width=17cm]{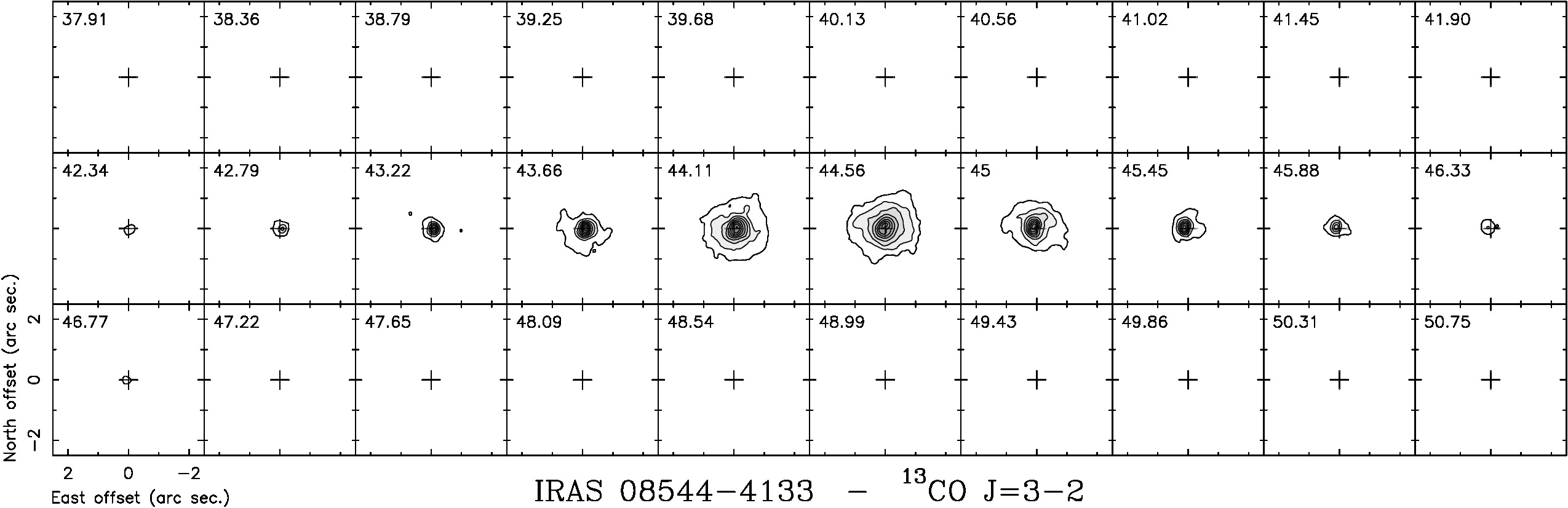}
      \caption{ALMA maps per velocity channel of \trece\ \jtd\ emission
        in \sou. The continuum has been also subtracted in this
        figure. The contours are shown as in Fig.\ 1: 0.02, 0.04,
        ... Jy beam$^{-1}$. The {\em LSR} velocities are indicated in each panel. }
         \label{}
\end{figure*}

\begin{figure*}
  
   \hspace{.3cm}
   \includegraphics[width=17.cm]{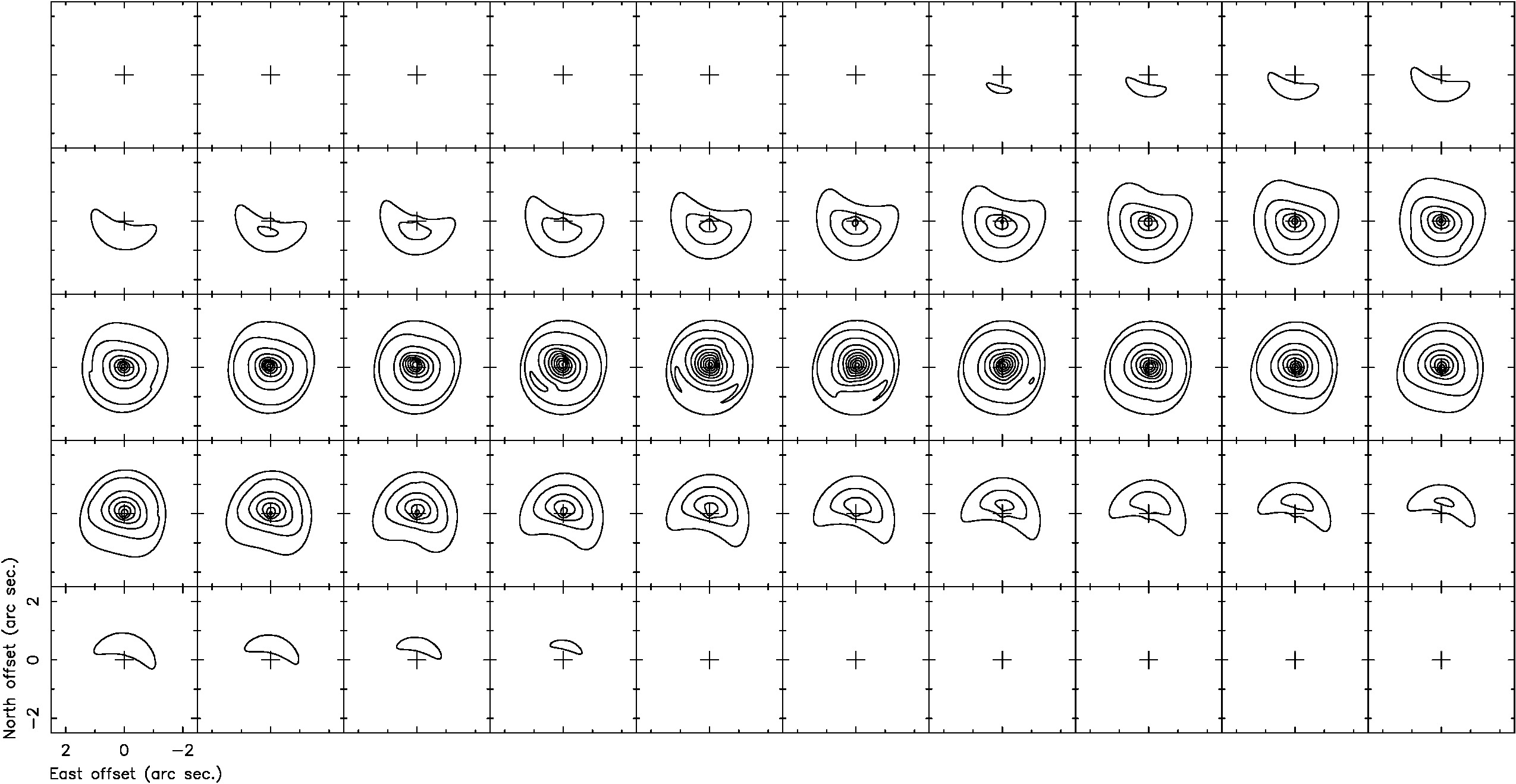}
    \caption{Synthetic maps per velocity channel from our best-fit model
      of \doce\ \jtd\ emission in \sou. These maps are comparable with Fig.\ 1; all scales and contours
        are the same as in that figure.}
         \label{}
   \end{figure*}

\section{Observations}

We present maps of \sou\ in the \doce\ and \trece\ \jtd\ lines,
$\lambda$ = 0.8 mm that were obtained with ALMA, band 7 receiver. A total of
five observing runs were performed: three between August 29 and 30,
2015, and two more in August 25 and September 3, 2016. The source was
observed for $\sim$ 37 min in each track. The observations were
obtained during ALMA Cycle 2. Thirty-four antennas were used with 
baselines ranging from 15 to 2483 m.

Data were calibrated with the CASA software package.  The quasars
J0538-4405, J0922-3959, and J0904-5735 were observed
for bandpass, flux, and phase calibrations.  Through a comparison of their
fluxes in the different runs, we obtained significant differences
between 2015 and 2016 observations, which is not exceptional in
measurements of quasar continuum flux. The fluxes of the calibrators
were found to change, respectively, from 886/372/464 mJy in 2015
to 1356/332/766 mJy in 2016. The flux calibration was however very
consistent between the observations performed in the same
year. Finally, by looking at final source data, in particular by comparing the
source continuum emission obtained from 2015 data and 2016
observations, the flux calibration was judged to be optimal and no
additional flux rescaling was applied.

After the data calibration, the rest of the analysis was made with
the GILDAS software package.  First, additional phase
self-calibration was performed using the compact continuum source as
reference. Image deconvolution was carried out with natural and
also robust weighting, which leads to channel maps with 0.19$\times$0.18
arcsec and 0.13$\times$0.10 arcsec (HPBW) resolutions, respectively. Various
CLEANing methods (Hogbom and SDI) were also used in the image synthesis
to best represent the different emission components. The SDI
method is known to be more adapted to represent the most extended
emission. The data here presented were obtained with
natural weighting and SDI method. 

The ALMA backend was set to achieve a spectral resolution of about 0.2
\kms; the data were delivered with a channel spacing of 0.11 \kms).  In
the data presentation selected for this paper, the resolution was
degraded to about 0.43 \kms\ to improve the S/N at high velocities. We
kept however the highest spectral resolution for the position-velocity
diagrams to give a better representation of the velocity structure of
the intense Keplerian disk. All velocity values in this paper are given
in the Local Standard of Rest ({\em LSR}) frame. 

By comparing  our data with APEX single-dish profiles (Bujarrabal et al. 2013a),
we  conclude that a small fraction of the flux, $<$ 20\%, was
filtered out in the maps of \doce\ \jtd. Anyway, we note that such moderate
difference is close to usual uncertainties in absolute flux
calibration.  In addition, both single-dish and integrated ALMA profile
shapes are very similar, confirming a low degree of lost flux.  The
\trece\ maps, with a less extended brightness distribution, are not
expected to show a significant flux loss.

Dust continuum emission was found to be not resolved and centered at RA
08:56:14.165 ~DEC $-$44:43:10.588, which is also the center of all the
maps here presented. Fitting of the continuum data in the uv-plane
yielded a total flux of $\sim$ 320 $\pm$ 1 mJy and an estimated size of
the emission region smaller than 0.1 arcsec. Such a continuum level was
subtracted from our maps to better show the weakest features, in
particular in the \trece\ position-velocity diagrams.

\section{Modeling of our ALMA maps}

The amount of data available for \sou\ and their quality are moderate, and these data
are not comparable to those obtained for the best-studied NIR-excess
post-AGB object, the Red Rectangle. We also lack information on the
nebula in general. Under these conditions, very detailed models, such as
those developed for the Red Rectangle, are not sensible and the
uncertainties for several derived parameters are not negligible, as
discussed below. Fortunately, the observational results in \sou\ are
not very different from those obtained for the Red Rectangle, AC Her,
89 Her, and IW Car (Bujarrabal et al.\ 2007, 2015, 2016, 2017). In view
of this, our models for \sou\ follow the general patterns found for
these objects.

We used codes that are very similar to those described in our previous works
(Bujarrabal et al.\ 2013b, 2015, 2017, ...).  As for similar objects,
all information we have on this nebula is compatible with the presence
of axial symmetry. We assume local-thermal-equilibrium (LTE) populations for the involved
rotational levels.
This is a reasonable assumption for low-$J$ CO
transitions in the dense material expected in our sources, $n$
\gsim\ 10$^4$ cm$^{-3}$, since their Einstein coefficients are then
smaller than the typical collisional rates; see further discussions in
Bujarrabal \& Alcolea (2013), Bujarrabal et al.\ (2016).  The use of
LTE may introduce some uncertainties (see below), but it significantly
simplifies the calculations and provides an easier interpretation of
the fitting parameters. For each considered model nebula, we assume a
shape for the nebula, constant molecular abundances, and distributions
of the local velocity dispersion, macroscopic velocity, density, and
temperature. With these ingredients it is possible to calculate the
absorption and emission coefficients of the two observed lines. These
are computed for a high number of projected velocities, according to
the actually observed channels, and for a high number of elemental
cells occupying the whole nebula, typically around 10$^6$ cells are
used in our calculations. The cell density in general oversamples the
central regions of the nebula, where the rotation velocity varies
faster. We then solved the standard radiative transfer equation in a
high number of directions pointing to the telescope (and for the set of
projected velocities), taking into account the assumed orientation of
the nebula axis with respect to the plane of the sky and to the
north. Typically, we solve the transfer equations following 10$^4$ to
10$^5$ rays, which traverse the cells in which the nebula has been
divided.  We get a predicted brightness distribution, as a function of
the coordinates (right ascension and declination offsets) and of the
projected velocity. This distribution is numerically convolved with the
interferometric clean beam and converted to units that are directly comparable
to the observational data, Rayleigh-Jeans-equivalent brightness
temperatures, or Jy/beam.

As mentioned, we took into account our results for better studied
sources to select possible nebula models. In any case, a high
number of different configurations were analyzed.

\subsection{Description of the best-fit model}

We finally adopted a model nebula as a good representation of the
source, based on its reasonable properties and comparing the
predictions with the observational data and other previous results; see
predicted maps per velocity channel and position-velocity diagram in
Figs.\ 3 and 5. Graphical representations of the main model parameters
are given in Figs.\ 6 and 7. The parameters describing the best-fit
model nebula, including those describing the main physical conditions,
are given in Table 1. In principle, we give our results for an assumed
distance $D$ = 1100 pc. We have seen in Sect.\ 1 that the distance
could be significantly lower; the effects of the assumed distance
on the derived parameters are discussed in Sect.\ 3.2, in which we consider
in particular the case of an alternative distance of 550 pc.

\begin{figure}
  
   \hspace{-0.cm}
   \includegraphics[width=8cm]{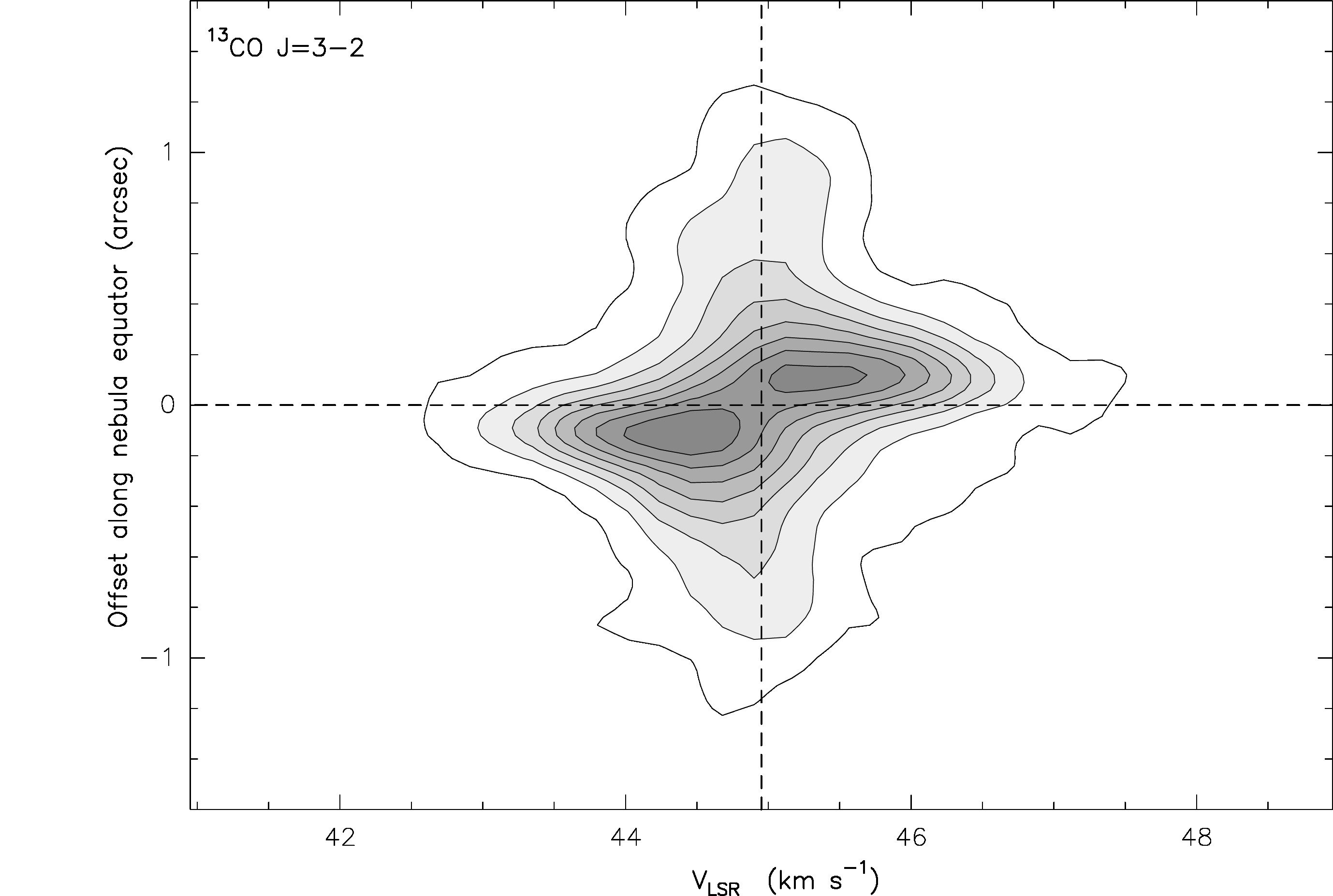}
      \caption{Position-velocity diagrams from our ALMA maps of
        \trece\ \jtd\ in \sou\ along the direction P.A.\ =
        75$^{\circ}$. Contours are the same as in the channel maps, but
        we used a higher spectral resolution to better show the
        velocity structure. The dashed lines show approximate
        centroids in velocity and position.}
         \label{}
   \end{figure}

\begin{figure}

  \hspace{.63cm}
   \includegraphics[width=7.35cm]{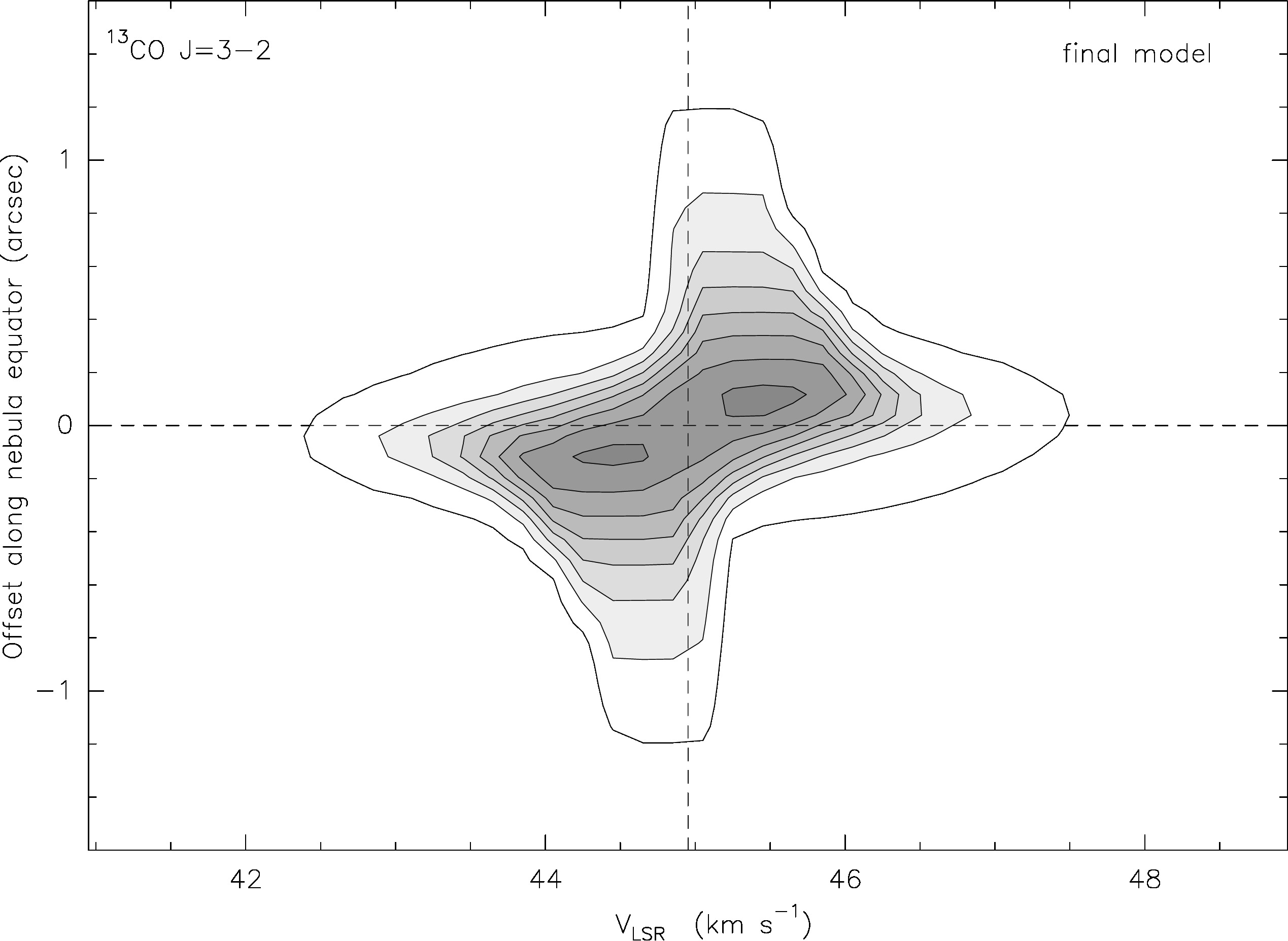}
      \caption{Synthetic position-velocity diagrams from our best-fit model
      of \trece\ \jtd\ emission in \sou, to be compared
        with Fig.\ 4; all scales and contours are the same as in that figure.}
         \label{}
         \end{figure}

\begin{table*}[bthp]
\caption{Physical conditions in the molecule-rich nebula, derived from
  our model fitting of the CO data, and assuming $D$ = 1100 pc. The values of the physical
  conditions depend on three geometrical parameters: the distance to
  the center, $r$, distance to the axis, $p$, and distance to the
  equator, $h$.  See
  Figs.\ 6, 7 for more details and  cartoons of the density, velocity, and temperature 
  distributions. } {\tiny
\begin{center}                                          
\begin{tabular}{|l|cc|cc|cc|}
\hline\hline
 & &  &  & & & \\ 
 & \multicolumn{2}{c|}{Inner disk} &  
\multicolumn{2}{c|}{Outer disk} & \multicolumn{2}{c|}{Outflow} \\
&   \multicolumn{2}{c|}{(pure Keplerian rotation)} &  
\multicolumn{2}{c|}{(subkeplerian rotation plus slow expansion)} & & \\
& & & & & & \\
{Parameter}  & {Law} & { Values} &   {Law} & {Values}  &   {Law} & {Values} \\ 
& & & & & & \\
\hline\hline
&  &  &  &  &  & \\
Radius & r $<$ $R_i$ & $R_i$ = 6 10$^{15}$ cm  & r $<$ $R_o$ &
 $R_o$ = 2 10$^{16}$ cm & p $<$ $R_{of}$ & $R_{of}$ = 2.5
10$^{16}$ cm \\
 (see Figs.\ 6, 7) &  &  &  &  &  & \\
\hline
&  &  &  &  &  & \\
Total width & $h$ $<$ $H_i$ & $H_i$ =  2.8 10$^{15}$ cm & $h$ $<$ $H_o$
& $H_o$ = 4  10$^{15}$ cm & $h$ $<$ $H_{of}$ &  $H_{of}$ = 3.4 10$^{16}$ cm  \\
(see Figs.\ 6, 7) & &  &  &  &  & \\
\hline
&  &  &  &  &  & \\
Temperature  &  $T$ = $T_o \left( \frac{5\, 10^{15} {\rm cm}}{p} \right)^{\alpha_T}$
&  $T_o$ = 36 K
 &   $T$ =  $T_o \left( \frac{10^{16} {\rm cm}}{p} \right)^{\alpha_T}$ &
 $T_o$ = 27 K  & 
  $T$ = $T_o \left( \frac{10^{16} {\rm cm}}{r} \right)^{\alpha_T}$ &
  $T_o$ = 50 K  \\ 
 & & $\alpha_T$ = 0.4 & & $\alpha_T$ = 1  &  & $\alpha_T$ = 0.7 \\
\hline
&  &  &  &  &  & \\
Density &  $n$ = $n_o \left( \frac{5\, 10^{15} {\rm cm}}{p} \right)^{\alpha_n}$
&   $n_o$ = 3.7  10$^6$ cm$^{-3}$ 
&  $n$ =  $n_o \left( \frac{10^{16} {\rm cm}}{p} \right)^{\alpha_n}$
&  $n$(10$^{16}$cm) = 9 10$^5$ cm$^{-3}$
&  $n$ =  $n_o \left( \frac{10^{16} {\rm cm}}{r} \right)^{\alpha_n}$
&   $n_o$ = 8.5 10$^4$ cm$^{-3}$   \\
  &  & $\alpha_n$ = 1 &  &  $\alpha_n$ = 1  &  &  $\alpha_n$ = 2.3 \\
\hline
&  &  &  &  &  & \\
Local velocity   & constant  & 0.1 \kms &  constant  & 0.1 \kms &  constant  &  2 \kms \\
dispersion   &  &  &  &  &  & \\
\hline
\end{tabular}
\begin{tabular}{|l|cc|l|}
\hline
 & & & \\
{Other parameters}  & {Law} & {Values}
 & comments \\ 
& & & \\
\hline\hline
 & & & \\
Axis inclination with respect to the plane of the sky & & 70$^\circ$ & from IR  and
CO data  \\
 & & & \\
\hline
 & & & \\
Axis inclination in the plane of the sky (PA) & & $-$15$^\circ$ & from
IR and CO data \\
 & & & \\
\hline
& & & \\
Distance  & & 1100 pc & various arguments (Sect.\ 1) \\
 & & & \\
\hline
 & & & \\
\doce\ relative abundance & ~~constant~~ & ~~1.5 10$^{-4}$~~  & this paper  \\
\trece\ relative abundance & ~~constant~~ & ~~1.5 10$^{-5}$~~  & this paper  \\
& & & \\
\hline\hline
\end{tabular}
\end{center}
}
\end{table*}

 In this best-fit model, we adopted a relative abundance with respect
  to the total number of particles $X$(\trece) $\sim$ 1.5 10$^{-5}$; as
  in most nebulae around evolved stars, we can assume that the dominant
  component of the low-excitation gas is H$_2$. This value of
  $X$(\trece) is similar to those adopted in our previous works to
  ease the comparison with previous results on this and similar
  objects.  Also following those works, we adopted an abundance ratio
\doce/\trece\ = 10. Those values lead to a reasonable fit of the data.

The disk is assumed to be formed of two components. In the inner component,
the rotation is purely Keplerian. The deduced velocity field
corresponds to a central (stellar) mass of about 1.8 \ms, reasonable
for this binary system (Sect.\ 1).  In the outer component, we assumed
sub-Keplerian rotation and the presence of a slow radial expansion,
which significantly helps to reproduce the data, as was also the
case in our studies of the Red Rectangle and IW Car (see discussions in
our previous works, Bujarrabal et al.\ 2005, 2013b, 2017). For the
sub-Keplerian velocity law we assumed angular momentum conservation.
Remarkably, a similar result was
independently found for L$_2$ Pup, the only AGB source in which a
rotating disk has been found to date, by Homan et al.\ (2017), who also
deduced sub-Keplerian rotation in the outer disk regions; in any case,
the size of the disk around L$_2$ Pup is much smaller than in our
objects.

We found that a small inner region of the disk must be devoid of
molecules in some way. It is necessary to adopt an assumption of this
kind to fit the high-velocity maps, but the angular resolution of our
data does not allow a proper description of this region (see
Sect.\ 3.2).  A similar result was also found in our previous works for the
Red Rectangle and IW Car. As in those papers, we assumed a progressive
decrease of the disk width in the inner regions, instead of a sudden
disappearance of the emitting gas. We recall that we cannot distinguish
from fitting the different options, and here we take a simple
law. An empty region at a smaller scale is also found in the IR VLTI maps
of hot dust emission in \sou\ (Hillen et al.\ 2016), which show
emission from a ring with a diameter of about 15 mas (1 -- 2 10$^{14}$
cm, compatible with the dust sublimation radius). Those observations
strongly select hot-dust inner regions and are probably able to trace
the very inner disk, at the vertex of the cone depicted in Figs.\ 6 and
7, where the emission of the hottest dust is prominent despite its
small radius and width.

In the disk, the density and temperature ($n$ and $T$) are assumed to
vary solely  with the distance to the axis $p$ and following potential
laws; see values of the parameters in Table 1. We note the laws are
not the same for both inner and outer disk components. The density
distribution is shown in Fig.\ 6. As in our previous works, we find
that the disk must show a very small local velocity dispersion to fit
the data; we assumed that this is the combination of the thermal
movements and a small local dispersion (microturbulence) of 0.1
\kms.

In the outflow, the density and temperature are assumed to vary with
the distance to the center, again following potential laws; see
parameters in Table 1. The expansion velocity is basically radial and has a minor component that is parallel to the equator. The local velocity
dispersion in the outflow is dominated by microturbulence with a
dispersion of 2 \kms. 

The total nebular mass derived from our fitting is $\sim$ 2 10$^{-2}$
\ms, about 90\% of which is placed in the disk. These values are
compatible with those found by Bujarrabal et al.\ (2013a, after
correcting the different assumed distance), although their treatment,
based only on single-dish \doce\ observations, was much more
uncertain. From the extent and velocity field of the expanding
envelope, we derived a typical time required to form it of about 1100
yr. From the disk/outflow mass ratio and assuming, as we deduced
for similar sources (Sect.\ 1), that the outflowing gas has been
expelled from the disk, we can estimate a disk lifetime of about 10000
yr. This value is comparable to that found for IW Car and the Red
Rectangle. This is just an estimate of the disk lifetime
scales, since the mass-ejection rate can vary with time.  We also
  stress that our outflow component is unbounded, since the escape velocity
  is $\sim$ 1.8 \kms\ at 1000 AU, a few times smaller than the outflow
  velocity.

As we discuss in Sect.\ 3.2, the derived parameters are affected by the
adopted distance, whose value is not well known (Sect.\ 1).  If we
assume a significantly shorter distance, as mentioned in the
Introduction, $\sim$ 550 pc, most model parameters change significantly
and \sou\ becomes a relatively small and low-mass post-AGB nebula.  The
total size would be $\sim$ 3.5 10$^{16}$ cm (lower than for the Red
Rectangle) and the total mass would become $\sim$ 6 10$^{-3}$ \ms\ (the
mass of the Red Rectangle is $\sim$ 10$^{-2}$ \ms). The central stellar
mass and the disk lifetime would become $\sim$ 0.9 \ms\ and $\sim$ 5000
yr, again lower than those of the Red Rectangle.  In view of the
parameters deduced for the stellar orbit, Sect.\ 1, a total mass of
about 1.8 \ms\ would be necessary to get a mass of the primary of about
0.5 \ms, a very reasonable value for a post-AGB star. A total mass of
$\sim$ 1 \ms, as deduced for 550 pc, would yield a primary mass of only
0.1 \ms. Therefore, this reasoning favors a long distance $\sim$ 1100
pc for \sou, which is taken as our standard value, but 
its validity depends on the relatively uncertain mass values.

 A distance of 1.1 kpc implies a luminosity $\sim$ 12000 \ls. Using
  the core-mass luminosity relation of Miller Bertolami (2016),
  this results in a post-AGB mass of about 0.65 M$_{\odot}$. But a
  larger primary mass requires a larger total mass (as deduced from the
  orbit mass function and inclination), and then still larger distance
  (to explain the observed disk rotation) and luminosity.  To be fully
  compatible with the stellar evolution calculations, we should adopt a
  distance $\sim$ 1.5 kpc, a primary mass $\sim$ 0.8 \ms, a total mass
  of 2.5 \ms, and a luminosity of about 2 10$^4$ \ls. However, this
  criterion is weak because of the poorly understood post-AGB evolution
  of the stars studied here (Sect.  1) and the difficulties in measuring  the involved parameters, widely mentioned in this paper. In
  particular, total stellar mass values of 2 -- 2.5 \ms\ are still
  compatible with a distance of 1.1 kpc and the observed Keplerian
  velocity field (within the uncertainties, Sect. 3.2).  The orbital
  parameters are also uncertain; in particular, the translation from
  observed orbital parameters to physical binary parameters is strongly
  dependent on the inclination.  An orbital inclination of 23$^\circ$,
  a change of just 3$^\circ$ with respect to our standard value, and a
  post-AGB mass of 0.65 \ms\ give a total stellar mass of 1.8 \ms\ 
  from the measured mass function. This is compatible with the
  gravitational mass obtained from the ALMA data and with evolution
  calculations for our distance value, 1.1 kpc. Moreover, a
  significantly larger distance would lead to very high values of the
  nebular mass and size compared with results for other similar
  objects (Sect. 1).  Accordingly, we think that the distance derived
  from the parallax measurements, 1.1 kpc, is a reasonable compromise that is
  compatible with all the available information.

 We derived the disk angular momentum by integrating the
  local momentum of the volume units for the adopted density and
  velocity laws. The uncertainties of this estimate are discussed in
  Sect.\ 3.2.  Particularly interesting is the distance dependence of
the disk angular momentum. For a distance of 550 pc, we derived a disk
angular momentum $J$ $\sim$ 1.6 \ms\,\kms\,AU. This value is
significantly smaller than that of the Red Rectangle ($\sim$ 9
\ms\,\kms\,AU), but still high for the low central mass deduced in that
case.  The strong dependence $J$ $\propto$ $D^3$ (Sect.\ 3.2) yields a
high value $J$ $\sim$ 13 \ms\ \kms\ AU for our best distance estimate,
$D$ = 1100 pc.   The angular
momentum of the stellar system at present is found to be $\sim$ 20
\ms\,\kms\,AU, comparable to that of the disk. The stellar
  parameters used for this estimate are given in Sect.\ 1; see
  more details and discussion in Van Winckel et al.\ (2009). Again an angle
between the orbit and sky planes of about 20$^\circ$ is assumed.

\subsection{Uncertainties in the derivation of the model parameters}

Some disk properties are not well determined from the observations,
mainly because of the relatively low angular resolution and the weak
dependence of the predictions on these properties.
In particular, the width of the
disk is not well determined because of the insufficient angular
resolution and significant inclination of the axis with respect to
the plane of the sky. The size and shape of the central region of the
disk is assumed to show a decreasing width, which is a result similar to that we
found in previously studied nebulae. The exact shape of these regions
is also difficult to measure, in fact, we take a typical diameter of 2
10$^{15}$ cm (for $D$ = 1100 pc) that is smaller than the resolution in linear 
units, 3 10$^{15}$ cm. An inner disk region with relatively low CO
emission clearly improves the quality of the modeling and is in fact
necessary to get a reasonable data fitting, but we must keep in mind
the uncertain structure of these very inner disk regions. On the
contrary, the diameter of the disk is well measured, since it is
basically given by its extent in the maps, that is, much larger than the
resolution.

The general structure of the wide outflow is also well determined from
the data. However, the details of the boundary shape are of course
uncertain, mainly in the farther regions of the X-shaped structure,
whose emission is weak. Because of the relatively large total extent,
comparable to the distance at which CO is photodissociated by the
interstellar UV field in expanding gas, it is probable that the actual
shape of the emitting region is rather due to CO photodissociation than
to a relatively sudden decrease of the density; for the relatively low
mass-loss rates characteristic of the outflows in our sources, see
Mamon et al.\ (1988), Groenewegen (2017).
Any attempt to deduce the size and shape of the CO-rich
gas from the dissociation theory is extremely difficult because of
the uncertainties in the path covered by each gas particle and the
unknown changes of the velocity with time.
We
think that the general nebula shape we propose in this paper, after
incorporating our previous experience with similar sources, is
realistic and could be transferred, with moderate scaling and changes,
to most of the sources of this kind.

The total mass of both nebular components is relatively well
constrained (for an assumed distance) because we match the total
intensity of optically thin emission: \trece\ emission for the disk and
inner outflow regions and \doce\ emission for the outermost
regions. The main source of uncertainty for the total mass comes from
the assumed value of the CO abundances, $X$(\doce, \trece). Because of
the LTE approach we used, there is a degeneracy between density and CO
abundance, such that predictions are identical for values of both
parameters keeping a constant product. The errors in the mass are,
therefore, inversely proportional to those of the
abundance. Fortunately, these abundances are not very uncertain,
showing a moderate variation between different studies. Following
discussions in our previous works, we estimate $X$(\trece) and
$X$(\doce) must  vary in the ranges 10$^{-5}$ -- 2
10$^{-5}$ and 10$^{-4}$ -- 2 10$^{-4}$, respectively. We so expect a moderate
uncertainty in the estimate of the molecule-rich gas mass, $\sim$
$\pm$50\%.   The degree of uncertainty in the estimate of the disk
  angular momentum is similar to that of the total mass (or just
  slightly larger), since the detected disk radius and velocity are
  relatively well measured. We cannot rule out the
  presence of outer regions that may remain undetected because of
  their low brightness or photodissociation. Therefore, the derived mass
  and momentum
  values may be lower limits corresponding to the actually detected
  gas.

The values of the temperature are relatively uncertain because we only
have data of \jtd\ emission. However, we note the high brightness of
optically thick emission (\doce\ line) from the disk and inner outflow,
$>$ 30 K, reaching values close to 100 K. These are comparable to the
high brightness found in the Red Rectangle, AC Her, and IW Car,
suggesting high temperatures similar to those we deduced in our
previous papers, over 100 K in central regions and decaying
outward.  Further details on the temperature distribution are
difficult to estimate.

The uncertainty in the average gas density in the disk can be
important, mainly because of the uncertain disk width, with variations of
the density inversely proportional to those of the width. The value
adopted for $X$ also affects the values deduced for the density, as
mentioned above.  The density is particularly difficult to estimate in
the outer regions of the outflow because of the low emission and more
uncertain temperature law there.  It is difficult to determine the
effect that the temperature uncertainty has on the density estimate,
but it is probably moderate. In principle, $n$ is appoximately
proportional to the assumed value of  $T$ for
temperatures much higher than the line excitation ($\sim$ 30 K), exact
LTE, and very low opacities. But the dependence is significantly lower
than linear and tends to vanish when the excitation temperature is
not very high or the optical depth is moderate.

The derived disk/outflow mass ratio and disk lifetime bear somewhat
stronger uncertainties than the independent mass values  because a
  change in the abundances affects the determinations of the 
  mass in both
  components in different ways. The uncertainty in the values of
$X$(CO) mentioned above leads to changes in the disk lifetimes between
5000 and 2 10$^4$ yr.

The velocity fields, both rotation and expansion, are relatively well
measured, since maps are obtained for well-known {\em LSR} velocities
and the nebula inclination is well constrained. Of course, the exact
direction of the velocity, i.e.,\ the inclination of the arrows in
Fig.\ 6, can significantly change. For instance, we cannot rule out a
purely radial velocity (as found for IW Car). In any case, we had
problems fitting the data with simple models that include radial
velocities; the assumption of a radial velocity field would probably
require small changes in the symmetry axis direction with the distance
to the center.
The central stellar mass depends on the square Keplerian velocity;
  therefore, even for small changes in the velocity of 15\%, we
  can expect variations in the stellar mass of about 30\%. The same
  applies for errors in the estimate of the inclination angle, since
  they affect the measurement of the Keplerian velocity modulus. We
  estimate that the stellar mass uncertainty is of about 40\%.

The distance of the object $D$ is very uncertain (Sect.\ 1), which
affects the other parameters. The
model nebula size must vary linearly with the distance to continue to fit
the data. The velocity field is not affected, provided that we scale
the velocity laws to the size of the nebula. The same rule holds for
the temperature. The density varies with $D^{-1}$, after scaling the
density law to the size of the nebula since the column density must be
conserved to yield the same optical depth in all lines of
sight. Therefore, the change in the total volume implies that the
total mass varies with $D^2$.  The variations of the rotation
velocities and distances mentioned above imply that the central stellar
mass varies proportionally to the assumed distance. The dependencies of
the velocity and size also lead to variations of the disk lifetime,
proportionally to $D$. The dependence of the disk angular momentum on
the distance is particularly strong, varying with the third power of
the assumed value, since the momentum of an elementary particle
rotating at a given velocity depends on its total mass and distance to
the axis. These dependence laws are basically the same
in all models of molecular line emission from AGB or post-AGB shells.

Finally, we note that some features of our observations are not well
reproduced by our model. Most of them are minor details, and probably
correspond to the actual complexity of the true nebula in comparison
with our very simple model. The most important discrepancy, in our
opinion, is the presence of a shift to relatively low {\em LSR} velocities
of the central maximum in the \doce\ maps. This would correspond to an
asymmetry in the emission between the gas that rotates approaching us,
which is brighter, and that receding from us. Since the effect is less
noticeable in \trece\ maps, the natural explanation is that the kinetic
temperature is higher in the disk edges at certain azimuthal angles. We
can speculate that the position of the central stars can lead to a
selective heating of certain regions of the disk. We do not try to
incorporate these phenomena in our nebula model, since their nature is
very uncertain and other complex effects could be present, such as
selective photodissociation and gas evaporation. The effects of such a selective
heating on the derived physical
parameters remains within the ranges discussed before.

\begin{figure}

  \hspace{-.0cm}
   \includegraphics[width=8.9cm]{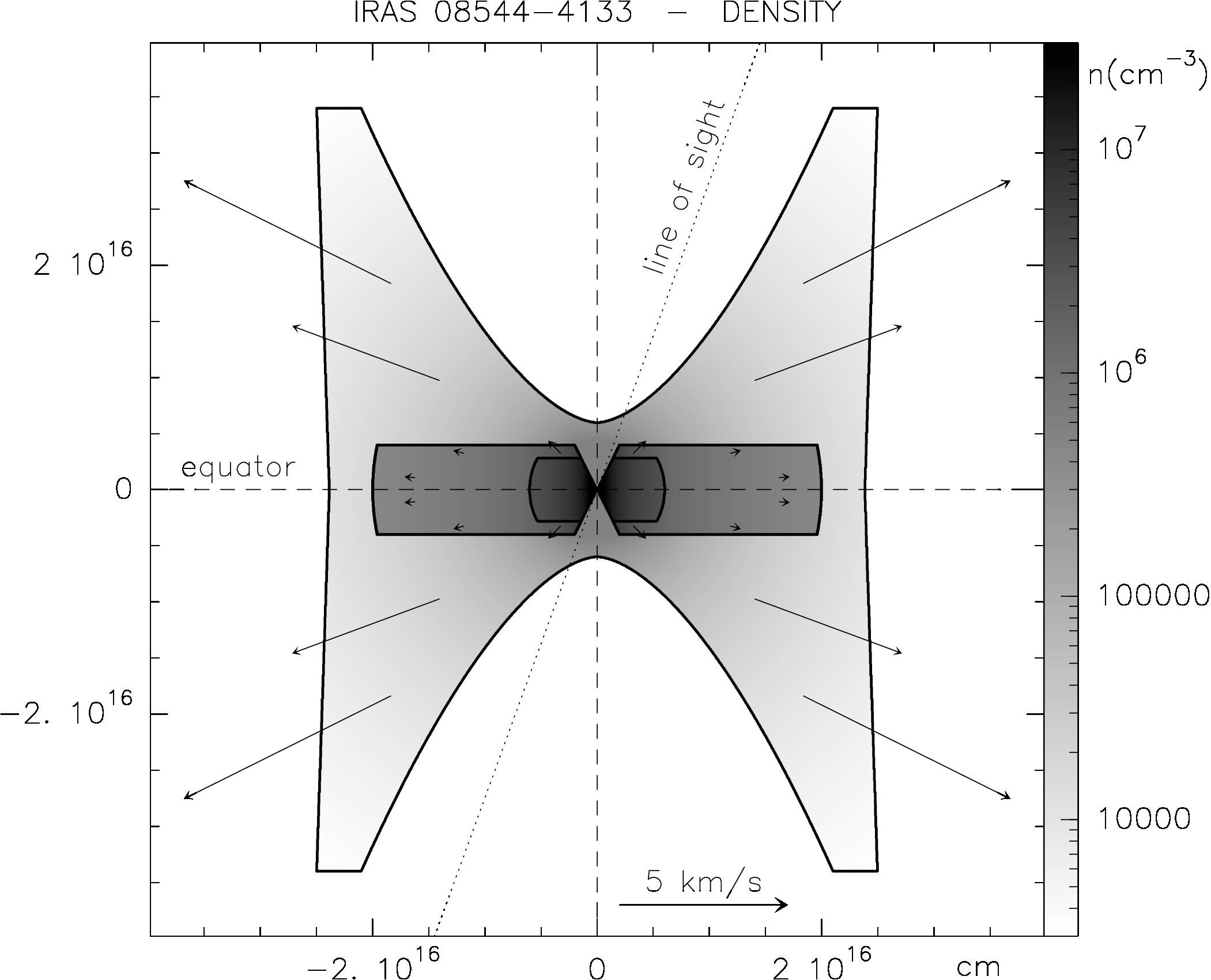}
      \caption{Density and velocity distributions in our best-fit
        model. The model is shown for $D$ = 1100 pc; the
        length scale would change proportionally for other 
        distance values. Only expansion velocities are shown because we
        represent a plane containing the symmetry axis; rotation
 is only present in the
        equatorial disk.}
         \label{}
         \end{figure}

\begin{figure}
  \hspace{-.0cm}
   \includegraphics[width=8.5cm]{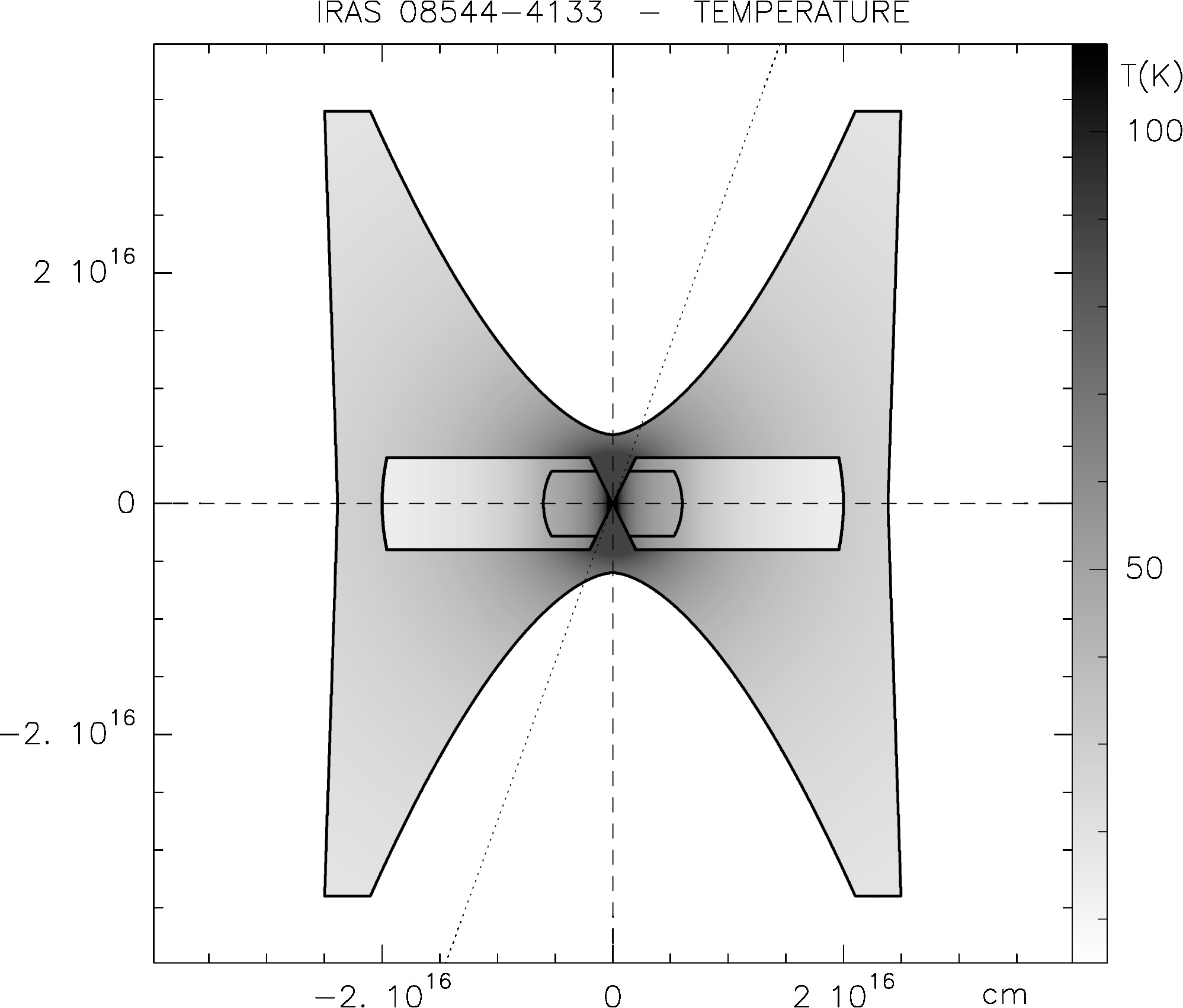}
   \caption{Temperature distribution in our best-fit model (for $D$ = 1100 pc). }
         \label{}
         \end{figure}

\section{Conclusions}

We present high-quality ALMA maps of \doce\ and \trece\ \jtd\ emission
from \sou\ (Sect.\ 2) and detailed modeling able to explain the main observational
features (Sect.\ 3). \sou\ belongs to a class of binary post-AGB
stars that are known to show indications of being surrounded by
material in rotation (Sect.\ 1), including a significant NIR excess
(supposed to be due to emission of hot dust) and peculiar CO line
profiles that are very similar to those expected from rotating
disks. The presence of rotating disks was previously confirmed by maps of the
velocity field in three of these  sources, the Red Rectangle, IW Car, and AC
Her. A component of gas in expansion, probably expelled from the disk,
 is also confirmed in five and probably
present in most of these objects  (Sect.\ 1). Both rotating and outflowing
components were only well detected in the Red Rectangle and IW Car. 
Our maps of \sou\ also show a
clearly composite nebula with a disk in rotation and gas in
expansion. The general properties of our maps and modeling of \sou\ are
remarkably similar to those found in better studied similar objects,
such as the Red Rectangle, which confirms our interpretation.

We analyze the CO emission by means of nebula models accounting for the
complex nature of the source.  From our model fitting, we derive the
main nebula parameters: shape and velocity field, density distribution
and total mass, and characteristic temperature (Sect. 3).  We
extensively discuss the uncertainties in the derivation of those
parameters in Sect.\ 3.2, in particular the dependence of the derived
properties on the distance of the star. We adopt a distance $D$ = 1100
pc, but we are aware of that this value is uncertain and also discuss
the case of a distance smaller by a factor 2 (Sect.\ 1), particularly
to compare our  conclusions with previous results. The mass of the nebula is
found to be $\sim$ 2 10$^{-2}$ \ms\ ($\sim$ 6 10$^{-3}$ \ms\ for $D$ =
550 pc), and about 90\% of the nebular material would be placed in the
disk. These values are compatible with those typically found in similar
sources, as well as with previous estimates for \sou\ from much less
complete data (Bujarrabal et al.\ 2013a).  The mass of the central
stellar system is derived from analysis of the rotation dynamics. We
find a central mass of about 1.8 \ms\ (0.9 \ms). The higher stellar
mass value is comparable to the result found for the Red Rectangle,
while 0.9 \ms\ is very similar to that of the IW Car. A high stellar
mass value is more compatible with the measured properties of the
binary system, which is composed of a very luminous post-AGB primary
with roughly 0.5 -- 0.8 \ms\ and a more massive secondary (see
Sects.\ 1, 3.1). In spite of the uncertainties, the need for a
relatively high total mass favors a distance of about 1100 pc for \sou.
The typical size of the nebula is $\sim$ 6 10$^{16}$ cm ($\sim$ 3
10$^{16}$ cm for $D$ = 550 pc). As for other well-studied objects, only
the central part of the disk is in purely Keplerian rotation; the
rotation of the outer disk is probably sub-Keplerian and a slow
expansion appears in it. A similar result was found in the Red
Rectangle and IW Car, and in the only AGB star in which a rotating disk
has been found, i.e., L$_2$ Pup (Homan et al.\ 2017)

It is remarkable that, for the low distance value we considered, i.e.,
$D$ = 550 pc, the nebula would show relatively low mass and size and the
total stellar mass would be relatively low compared to those of the Red
Rectangle (the best studied object of this class), but  closer to
the properties of IW Car. However, for our best distance value, $D$
$\sim$ 1.1 kpc, \sou\ is slightly larger and more massive than
the Red Rectangle.  We propose that the Red Rectangle and \sou\ are
relatively similar objects.

 The angular momentum found in the disk (for $D$ = 1.1 kpc) reaches
  a high value $J$ $\sim$ 13 \ms\ \kms\ AU, which is comparable to that found
  for the binary system at present ($\sim$ 20 \ms\ \kms\ AU, see
  Sects.\ 3.1 and 1). In our case, it is expected that the disk
  angular momentum comes from the binary system because the gas is
  ejected with negligible rotation. Therefore, and keeping in mind the
  uncertainties that affect these measurements, we deduce that the
  binary angular momentum was $\sim$ 33 \ms\ \kms\ AU in the
  past. Since in a binary star $J$ is basically proportional to the
  square root of the orbit size, we conclude that the distance between
  the stars has significantly decreased, by a factor \gsim\ 2, owing to
  the transfer of angular momentum to form the disk. The orbit was
  probably larger than an AGB star, but not much larger, which allowed
  a significant momentum transfer. The moderate orbit size change 
  indicates that the system certainly had momentum
  enough to explain the disk rotation, while maintaining a size of some 
  astronomical units during the whole process. We hope that these
  results serve to better understand the evolution of binary stars
  in the presence of dense circumstellar material and as a comparison
  with theoretical studies of the transfer of angular momentum to
  circumbinary disks (Chen et al.\ 2017, Akashi \& Soker 2008,
  Dosopoulou \& Kalogera 2016, etc).

We conclude that these NIR-excess post-AGB objects systematically show
composite nebula, which contain relatively extended disks in rotation, plus
gas in slow expansion that is probably escaping from the disk. The
total mass of such nebulae is small compared with those of most PNe
and pPNe (Sect.\ 1), i.e., $<$ 10$^{-1}$ \ms\ and often $\sim$ 10$^{-2}$ \ms.
 The mass of the outflow is several times lower than that of the disk
in well-studied cases.  \sou\ is the third object in which these nebular
structure and dynamics are well established and the main properties of
both components are described.

\begin{acknowledgements}
This work has been supported by the Spanish MINECO (grants
AYA2012-32032, FIS2012-32096, and AYA2016-78994-P), and by the European
Research Council (ERC Grant 610256: NANOCOSMOS).  We used the SIMBAD
database to check some properties of the source. We are grateful to the
referee of this paper, Dr.\ O.\ de Marco, for her constructive
comments. This paper makes use of the following ALMA data:
ADS/JAO.ALMA\#2013.1.00338.S. ALMA is a partnership of ESO (representing
its member states), NSF (USA) and NINS (Japan), together with NRC
(Canada), MOST and ASIAA (Taiwan), and KASI (Republic of Korea), in
cooperation with the Republic of Chile. The Joint ALMA Observatory is
operated by ESO, AUI/NRAO and NAOJ. 
\end{acknowledgements}


{}

\end{document}